\documentclass[aps,prx,10pt,notitlepage]{revtex4-2}

\usepackage{amsmath,amssymb,amsfonts,mathrsfs,graphicx,upgreek,gensymb}
\usepackage{color}

\begin{document}

\title{Physical characteristics of mixed-species swarming colonies}

\author{Ajesh Jose}

\affiliation{Zuckerberg Institute for Water Research, The Jacob Blaustein Institutes for Desert Research, Ben-Gurion University of the Negev, Sede Boqer Campus 84990, Midreshet Ben-Gurion, Israel}

\author{Gil Ariel}

\affiliation{Department of Mathematics, Bar-Ilan University, 52000
  Ramat Gan, Israel}

\author{Avraham Be'er}

\affiliation{Zuckerberg Institute for Water Research, The Jacob Blaustein Institutes for Desert Research, Ben-Gurion University of the Negev, Sede Boqer Campus 84990, Midreshet Ben-Gurion, Israel}
\affiliation{Department of Physics, Ben-Gurion University of the Negev 84105, Beer-Sheva, Israel}

\begin{abstract}

In nature, bacterial collectives typically consist of multiple species, which are interacting both biochemically and physically.
Nonetheless, past studies on the physical properties of swarming bacteria were focused on axenic (single species) populations.
In bacterial swarming, intricate interactions between the individuals lead to clusters, rapid jets and vortices that depend on cell characteristics such as speed and length.
In this work, we show the first results of rapidly swarming mixed-species populations of {\sl Bacillus subtilis} and {\sl Serratia marcescens}, two model swarm species that are known to swarm well in axenic situations. In mixed liquid cultures, both species have higher reproduction rates.
We show that the mixed population swarms together well and that the fraction between the species determines all dynamical scales - from the microscopic (e.g. speed distribution), mesoscopic (vortex size) and macroscopic (colony size structure). 
Understanding mixed-species swarms is essential for a comprehensive understanding of the bacterial swarming phenomenon and its biological and evolutionary implications.

\end{abstract}

\maketitle

%\clearpage

\section{Introduction}
\label{sec_intro}

Bacterial swarming is an intricate mode of cellular motion in which flagellated bacteria collectively migrate in groups on surfaces \cite{ref1,ref2,ref3,ref4,ref5,ref6,ref7,ref8,ref9,ref10,ref11,ref12}. The swarming phenomenon is initiated in the cells by alternation of some of their physical and biological characteristics, for instance changes in cell length (aspect ratio), secretion of surfactants, synthesis of extra flagellin that leads to the growth of many flagella, and expression of other key proteins \cite{ref2,ref5,ref13,ref14,ref15,ref16}. This indicates that the transition into swarm cells requires energy and resources. Accordingly, several evolutionary advantages have been suggested to explain why bacteria follow this collective movement pattern. For example, it was found that swarming enhances expansion rates and dissemination of individuals \cite{ref17,ref18,ref19}, it promotes access to nutrients, as well as increased survival rates under harsh condition such as lack of moisture \cite{ref2}. In addition, swarm cells have shown a higher resistance to antibiotics compared to planktonic (free-swimming) cells \cite{ref20}. Bacterial swarming is not species specific, and a large variety of bacteria may exhibit this type of dynamics in-vitro. Some of the known swarming species are {\sl Escherichia coli} \cite{ref12}, {\sl Bacillus subtilis} \cite{ref8}, {\sl Serratia marcescens} \cite{ref21,ref22}, {\sl Proteus mirabalis} \cite{ref13}, {\sl Paenibacillus dendritiformis} \cite{ref23,ref24}, {\sl Pseudomonas aeruginosa} \cite{ref3,ref25} and {\sl Salmonella typhimurium} \cite{ref26}.

Most of past swarm studies were focused on axenic (single species) cultures. While heterogeneous swarms consisting of several strains of the same species were recently studied \cite{ref27,ref28}, multispecies swarms are much more complex and were rarely considered \cite{ref29}. In light of the fact that natural niches typically consist of multiple species, which is true for most species in the animal kingdom \cite{ref29a}. 

In this work, we mix two swarming species, {\sl B. subtilis} and {\sl S. marcescens}, and study the physical characteristics of a joint swarm colony. The two species exhibit different swarm behaviors if grown alone but alter their behavior if mixed. We focus on the dynamics and structure of mixed colonies as a function of the partial ratio between the two species and show that despite a well-mixed spatially homogeneous inoculum, these species partially separate to form spatially-heterogeneous spatial distributions. Overall, both species seem to benefit from the mixing, which is manifested in increased swarming speeds and longer correlations. Our study is different from those obtained for immotile mixed cultures, e.g., biofilms \cite{ref30,ref31}, because ours involves intense and rapid dynamics. Thus, our results open a door for new studies of migrating mixed populations.

From a physical perspective, bacterial swarm studies may be divided into two regimes: macroscopic, e.g., \cite{ref32} namely how a swarm colony expands on a semi-solid surface, and microscopic, e.g., \cite{ref4}, describing how cells within the swarm move and interact with each other. The two are not necessarily correlated, as the microscopic dynamics does not directly determine the macroscopic expansion rate (or vice versa) \cite{ref10,ref23}. 

In {\sl B. subtilis} that grow on a soft agar, a swarm colony typically has the form of a quasi-circular envelope-like pattern that expands outwards. The exact structure and the expansion rate depend on the available nutrients, but mainly on the availability of water (agar concentration) and the surface tension \cite{ref33,ref34}. On harder agar concentrations, {\sl B. subtilis} form a branched colony and the swarm dynamics is significantly reduced. While {\sl B. subtilis} tend to form thin-layer colonies, {\sl S. marcescens} form a much thicker structure with hundreds of layers of cells rigorously moving one on top of the other, in a synchronized fashion \cite{ref21}. The structure of {\sl S. marcescens} colonies is less sensitive to agar concentrations and their thick swarm dynamics is exhibited under a variety of environments. 

On the microscopic scale, swarming is characterized by dynamic clusters of cells that move in coherent whirls and jets \cite{ref4,ref6,ref10,ref11,ref21,ref23}. Recent works have shown that the two main mechanisms involved in swarming are long-range hydrodynamic forces \cite{ref35,ref36}, where the viscous media in which the cells swim is influencing their dynamics, and short-range repulsive steric forces between the rod-shaped cells that builds temporary rafts, or clusters \cite{ref11}. These past findings suggest that heterogeneous swarms with populations differing in one or more characteristics may form a much more intricate dynamics because the interactions between the individuals would be complex. Indeed, recent work on mixed populations with strains of the same species that differ in length or speed have shown that the dynamics of the group is strongly affected by the fraction of each sub population \cite{ref27}. Also, in motility defective cells that were generated due to exposure to antibiotics, the sub populations self-segregated into immobile clusters that effectively increased the average mobility of the group \cite{ref28,ref37,ref38}.

Below, we describe the experimental methods and main results of three aspects of the swarm: (i) The macroscopic structure of the colony, (ii) The microscopic swarm dynamics, and (iii) Biochemical interactions. To the best of our knowledge, these complementary points of view provide a first broad description of mixed-species swarms.

\section{Colony structure}
\label{sec_colony}

\subsection{Methods}

The bacterial species used in this study were {\sl Bacillus subtilis} NCIB 3610 “wild-type” and {\sl Serratia marcescens} 274 “wild-type”. Both species were fluorescently labeled using different fluorescent proteins to distinguish between them in a mix. {\sl B. subtilis} was labelled red (pAE1222-LacA-Pveg-R0\_mKate, amp), and {\sl S. marcescens} was inserted with a green fluorescent plasmid (pBEST deGFP Nhis). In both, the fluorescence was stable throughout the experiment and did not show any cost of carriage or photobleaching. Antibiotics, at a concentration of 100 $\upmu$g/ml was added to frozen stocks and to hard-agar (2\% and 25 g/l Luria Bertani (LB)) isolation plates; streptomycin to {\sl B. subtilis} and carbenicillin to {\sl S. marcescens}. Experiments were performed without additions of antibiotics and fluorescence was completely stable. 

Isolated colonies from hard-agar plates were cultured separately (each species on its own) in a 2 ml LB liquid medium (15 ml tubes) and incubated at 200 rpm and 30$\degree$C for 17 hours. The overnight cultures were mixed at different ratios in small vials and vortexed for a few seconds before inoculation to the swarm plates; a small 4 $\upmu$l drop at the center of a swarm plate. Swarm plates are standard 8.8 cm Petri-dishes filled with 15 ml of molten 0.6\% agar supplemented with LB. Swarm plates were prepared 20 hours prior to inoculation and aged at 22$\degree$C and 40\% relative humidity (RH), then 8 mins in the flow hood. After inoculation, swarm plates were incubated at 30$\degree$C and 95\% RH until the diameter of the swarm colony was 4 cm, as observed under an optical microscope. These particular conditions were chosen to equally meet the favorable growth conditions of each species.
 
An epifluorescence optical microscope Axio Zen 16 (Zeiss) was used to follow the expansion and dissemination of each species across the colony. The field of view in the microscope is large, up to about 1”$\times$1”, but individual fluorescence cells are still resolved. Both species were tracked by two different channels, {\sl B. subtilis} in a red channel and {\sl S. marcescens} in a yellow channel (a yellow channel was used instead of a green one as the excitation light of the GFP setup destroys cell motility). These two channels were superimposed. For maximal resolution, tiles-stitching of sufficiently large images (37$\times$) was performed. 

Fluorescence intensity was calculated from each image tile to estimate the thickness of the colony, which is quantified by comparing the fluorescence intensity from the virgin agar surface with a single- or double-layer colony. Virgin agar composed of LB showed some weak excitation for fluorescence resulting in a threshold fluorescence intensity value across the colony, which was reduced from the fluorescence intensity value obtained from the colony's interior. Compared to single layered colonies, two-layered colonies showed, on average, a double excess fluorescence intensity. This suggests that, at low magnifications, where the focal plane is relatively deep, fluorescence intensity increases approximately linearly with the number of layers in the colony. This observation is used to estimate the depth of the colony. Note that {\sl S. marcescens} secrete a red pigment, which caused self-fluorescence in the red channel. However, the intensity of this fluorescence is negligible compared to the red fluorescence of the {\sl B. subtilis}. 

High resolution off-focus fluorescence imaging was used to study the distribution of species in the colony volume \cite{ref21}. Here, cells located at the bottom of the colony near the agar surface appear in focus, whereas cells located at higher layers exhibit a diffraction ring, with a diameter that is proportional to their height from the focal plane (the agar surface). Elongated cells exhibit an oval shaped “ring”, where only the smaller diameter is considered for estimating the height. This technique was previously used to understand the 3D architecture of bacterial swarms \cite{ref21}. In the current study, we used an Optosplit II, Andor device, hooked to a Zeiss Axio Imager Z2 microscope and a 63$\times$ LD lens. The system splits a dually excited image (Ex 59026x, beam splitter 69008bs, and Em 535/30; 632/60) on a NEO camera (900$\times$1800 and 50 fps) in order to generate two simultaneous but separate fields of view, i.e., green (called green but may be considered yellow) and red, that are then merged again following post processing. In general, the {\sl B. subtilis} (labeled red) appear on the left panel, and the {\sl S. marcescens} (labeled green) appear on the right panel. In the off-focus mode, each of the labeled (red or green) populations was diluted within their WT non-labeled strains, at a ratio of approximately 10\%. For instance, in a 50:50 mix, this yielded $\sim$45\% white (non-labeled) {\sl B. subtilis}, $\sim$45\% white (non-labeled) {\sl S. marcescens}, $\sim$5\% red {\sl B. subtilis} and $\sim$5\% green {\sl S. marcescens}. Using this method, the diffraction rings are sufficiently sparse to be tracked, creating a 3D map of the colony for both species. 

\subsection{Results}

Experiments were repeated with different initial partial ratios between the species. Figure 1 shows an example for such a mixed colony with initial ratio of 75\% {\sl B. subtilis} and 25\% {\sl S. marcescens} (termed “B75” or “S25”). Figures 1a-b show a top-view two-dimensional fluorescence intensity image suggesting that the density of cells, and their position on the agar, differ between the species. The two panels show the same region at the same time. While {\sl B. subtilis} (red channel) grew faster and expanded outwards, {\sl S. marcescens} (green channel) expanded slower. Also, although the initial density of the two species was comparable (75-25 favoring {\sl B. subtilis}), the {\sl B. subtilis} formed an apparent sparse monolayer structure with a thicker narrow perimeter, while {\sl S. marcescens} are much denser, especially in the inner regions of the colony. It is therefore an indication that in the mixed colony the reproduction rate of the {\sl B. subtilis} is much slower than the one obtained for {\sl S. marcescens}. Figures 1c-e show a closer look of a mixed colony at different positions along its radius. It can be seen that the fluorescence intensity of each of the species changes with the distance from colony center. Note that the green stripes are found in axenic {\sl S. marcescens} colonies as well; bright stripes are thicker regions. 

Figures 2a-b summarize the fluorescence intensity of each species, from a top-view, for a variety of initial ratios between the species. The intensity of the two channels is normalized so that a single cell layer has the value of 1 in arbitrary units (AU); fluorescence intensity at this magnification is relatively constant along the entire focal plane that is several 10’s of $\upmu$m wide. While {\sl B. subtilis} tends to exhibit fluorescence values of the order of 2 (corresponding to 2 layers of cells), values of {\sl S. marcescens} are 10-fold larger. We stress that the difference in intensity between species is not due to differences in single-cell intensity but due to the fact that the {\sl S. marcescens} grow much thicker structures. The data presented in Fig. 2 shows the accumulated and averaged number of layers of fluorescence cells for each species in the mixed colony at a variety of distances from the center of the colony. 

In order to obtain the structure of the colonies, each one of the species was observed by a high-resolution fluorescence microscopy operated at an off-focus mode. Using the off-focus mode, most cells are not in focus and the diameter of the diffraction ring depends monotonically on the height of a cell from the agar surface – see Fig. 2c and Movie S1. Thus, the density of the moving cells in the volume of the colony may be calculated separately for each species in the mixed colonies. The results show that the 3D structure of the mixed colonies is very similar to the fluorescence density of the {\sl S. marcescens} (Fig. 2b), which occupy the volume almost uniformly (except for the very top; see also \cite{ref21} for more details), with {\sl B. subtilis} migrating within the volume. Thus, the single/dual layer(s) of fluorescence in {\sl B. subtilis} cells seen in Fig. 2a, may be misleading, as this layer is mixed inside the large volume of the entire colony, occupying mostly the two lower thirds of the local height of the colony. An example of a cross section of the structure of a mixed colony, and the distribution of the two species in it, can be found in Fig. 2d. All in all, the above results show that the two species, although they swarm jointly, mix along the z-axis but do not mix well on the $xy$ plane (the agar plane) and rather form a heterogeneous structure.

\section{Swarm dynamics}

\subsection{Methods}

The cells of each of the species move in a 3D space. However, 3D trajectories of all cells are difficult to track when using the off-focus fluorescence mode. Therefore, the collective motion of the cells was studied using optical flow (OF), where the two species are not distinguished, and their collective motion is studied instead. The same optical microscope (Zeiss Axio Imager Z2) was used to record microscopic swarm dynamics of mixed populations in a phase-contrast mode (63$\times$). The movies were captured at 100 fps and 1024$\times$1024 pixels for 15 seconds. These movies were captured 100 $\upmu$m from the edge of the colony, when the colony diameter was 4 cm. Movies streamed directly to the hard disk, resulting in a sequence of 1,500 images per experiment. Recorded movies were analyzed using MATLAB with the Horn Schunk method to obtain the optical flow (OF) between each consecutive frames \cite{ref39}.

\subsection{Results}

Figures 3a-b show an example snapshot from the region of interest: a top-view of the cells, and their velocity field. Data was collected from the most active region of the colony, as explained above. The microscope was fixed to focus on a plane located at half the height of the colony; See also Movies S2-4. Figures 3c-e show the average microscopic speed of the mixed cells, the fourth moment (kurtosis) of the distribution of velocities, and the correlation length of the velocity field (an example of a single experiment from which we derive the data is shown in Fig. S1). The latter is defined as the distance $r$ in which the value of the correlation function $C(r)=\left< v(x,y,s)v(x+\Delta x,y+ \Delta y,s) \right>-\left<v(x,y,s)\right>^2$ reduces by a factor of $e$, where $v(x,y,s)$ is the flow field obtained at position $x$, $y$ and time $s$, and angular brackets denote averaging with respect to $x$, $y$, and $s$ such that $\Delta x^2+\Delta y^2=r^2$. As can be seen, mixing the two species qualitatively changed the overall collective dynamics compared to the axenic cases. First, higher speeds were obtained for the mixed colonies compared to single species colonies of either {\sl B. subtilis} or {\sl S. marcescens}, indicating a strong, positive interaction between species. Second, the velocity distribution of the mixed populations transitioned from non-Gaussian with kurtosis$>4$ for axenic colonies, to a Gaussian, with kurtosis=3, indicating that the collective dynamics changes depending on the species composition. Third, the correlation length of the velocity field is much larger for mixed colonies, indicating larger vortices. On the other hand, while the macroscopic expansion speed of the mixed colonies is also mix-dependent, the dependence is monotonic, with higher speeds for larger ratios of {\sl B. subtilis}, suggesting that microscopic dynamics does not influence the expansion rate of the colony. Figure 3f shows an example of the macroscopic expansion of the farther front of the mixed colonies. The inset shows that the inner {\sl S. marcescens} front diameter, $d_s$, (as can be seen e.g., in Fig. 1b) grows faster as the fraction of {\sl B. subtilis}, suggesting that the {\sl S. marcescens} uses the fast-expanding {\sl B. subtilis} for a better dissemination.

\section{Biochemical interactions}

While the main scope of this work is the physics of the mixed cells, a large variety of biochemical interactions may occur if two bacterial species are mixed and grown as a single colony. Such interactions may lead to repulsion between the species, inhibition, enhancement or reduction of growth, or attraction. They may cause nothing, or lead to a complete extinction of one, or both of the species. In order to follow potential biochemical interactions between the two species, cultures were mixed and grown in liquid media with shaking. The latter reduces the effect of physical interactions such as cell-cell contact forces, hydrodynamic interactions, flow in thin layers, motility, and cell-surface interactions.

\subsection{Methods}

Separately for each species, colonies were cultured without antibiotics, in 2 ml LB liquid medium, and incubated at 200 rpm and 30$\degree$C overnight until they reached their stationary phase. Fresh cultures were prepared and incubated at 200 rpm and 30$\degree$C for 4 hours for {\sl B. subtilis} (OD=0.85 at 600 nm) and 2 hours for {\sl S. marcescens} (OD=1.1 at 600 nm). With this protocol, the number of bacterial cells of each species is approximately the same. After incubation, both bacterial species were mixed at different ratios. From the mixed cultures, 50 $\upmu$l was inoculated into 2 ml LB medium and incubated for 4 hours. Then, the mixed cultures were serially diluted and inoculated by spreading evenly on counting plates. Counting plates are standard 8.8 cm Petri-dishes filled with 20 ml LB and 2\% agar, prepared 20 hours prior to inoculation and aged at 20$\degree$C and 40\% relative humidity. These plates were incubated at 30$\degree$C for 16 hours. Colonies were differentiated by color; see Fig. S2. 

For supernatant experiments, 50 $\upmu$l of the fresh culture of each species was inoculated into 2 ml of the other species’ supernatant. Supernatant was extracted by filter sterilization of fresh (4 h) culture of each species. A 0.22-micron filter was used for the filter sterilization. These cultures were incubated for 4 hours followed by serial dilution and inoculation on the counting plates. 

For U-tube chamber experiments, 200 $\upmu$l of the fresh culture of each species was inoculated into each chamber containing 8 ml LB. Each chamber was separated from the other using a 0.45 micron amphoteric polyamide membrane (Schleicher and Schuell), see Fig. S3. These chambers were incubated for 4 hours while shaking (same protocol) followed by serial dilution and inoculation on the counting plates. The number of grown cells in each of the species is an indicator of the growth rate within the mixed liquid medium.

\subsection{Results}

The results indicate that the reproduction rate of both species increases (the doubling time is shortened) when mixed. Figure 4a shows that while the doubling time of {\sl B. subtilis} is 58 min when grown alone, it reduces to 44 min if the initial ratio between the species is 50\%. Similar results are found for {\sl S. marcescens}, which reduced the doubling time from 45 min while alone to 32 min in an initial 50:50 mix. These results suggest that the mix causes a faster reproduction of {\sl S. marcescens}, which may explain why the colonies are indeed richer in {\sl S. marcescens}, even though the initial ratio was in favor of the {\sl B. subtilis}. 

In a second set of liquid experiments, the two species were grown separately in the other species’ supernatant (Fig. 4b). It was found that when the other species is not present in the growth, but only its supernatant (meaning that there is no interaction but only response), {\sl B. subtilis} do not show a significant change in growth rate with respect to the control. {\sl S. marcescens} also show a minor reduction in the growth rate. This suggests that the dramatic increase in growth rate during mixed swarming is due to real cross-species interactions that requires action and response in real time, so that the presence of the other species during the growth is essential. 

In a third set of liquid experiments, the two species were grown in the same bulk liquid, but separated by a membrane. While the cells are unable to cross the membrane, signaling molecules, peptides and other materials that at smaller than the 0.45 micron pores may pass. The results, depicted in Fig. 4c, indicate that, while the doubling time of {\sl S. marcescens} shortened significantly (cells reproduce faster), the growth rate of {\sl B. subtilis} did not change. This indicates that {\sl B. subtilis} need to be in direct contact, or at least in very short proximity, to {\sl S. marcescens} in order to reproduce faster. 

The summary of the above results suggests that both species will benefit in a mixed culture only if they are in direct contact. The swarm assay shows a spatial macroscopic separation between the species at the front. Such a separation prevents the cells from being in contact and thus, the {\sl B. subtilis} located (nearly alone) at the front do not reproduce fast enough and their total number remains small. 

Lastly, we have conducted chemotaxis assays for swarming experiments by letting axenic colonies grow in short proximity (see an example in Fig. 4d). The colonies of both species did not show a preferred direction of expansion, suggesting that the two species do not exhibit attractive or repulsive chemotaxis during swarming.

\section{Discussion}

Mixed species were previously studied in several qualitative works, but did focus on the dynamics. In \cite{ref40}, Ingham et al. showed that motile bacteria carried an immotile fungus. The species disseminated, and the immotile fungus formed bridges over cracks in the medium to allow the joint swarm to expand. Rosenberg et al. \cite{ref30}, showed an extreme example of acute competition interactions. They studied interspecies interactions between biofilms of {\sl Bacillus simplex} and {\sl B. subtilis}. In their experiments, proximity between the biofilms caused {\sl B. subtilis} to engulf the competing {\sl B. simplex} colony. Upon interaction, {\sl B. subtilis} secreted surfactin and cannibalism toxins at concentrations that were inert to {\sl B. subtilis}. This eliminated the {\sl B. simplex} colony. However, cannibalism occurred on significantly slower timescales compared swarm dynamics, in which motion and mixing are important. Mashburn et al \cite{ref41} mixed {\sl P. aeruginosa} and {\sl Staphylococcus aureus}. They proposed a model where {\sl P. aeruginosa} lyses {\sl S. aureus} and uses released iron for growth in low-iron environments. This work however shows only the growth rate and did not consider the microscopic physical dynamics of the species. Finkelshtein et al., \cite{ref42} have shown that a motile antibiotic-sensitive bacterial species, {\sl Paenibacillus vortex}, carried an immotile antibiotic-resistant species and used its resistance to protect itself in further toxic environments. This involves interactions of cooperation between different bacteria, when one species provides an enzyme that detoxifies the antibiotic, while the other ({\sl P. vortex}) moves itself and transports the antibiotic-resistant species. This 2-species cooperative interaction was modeled in \cite{ref42a}. Again, their work did not study the microscopic dynamics. 

In contrast, the present work describes experimental results describing the structure and dynamics of mixed swarm colonies composed of the two species {\sl B. subtilis} and {\sl S. marcescens} at different initial ratios between them. First, our study shows the so-called simple, yet previously unknown result, that mixtures of different bacterial species can indeed exhibit swarming. Specifically, we show that at the macroscopic scale, the swarm morphology of the mixed colony follows the structural colonial shape of the dominant (in number) species (Fig. 2). Because the {\sl S. marcescens} reproduces much faster, most initial ratios will eventually yield thick colonies with mostly {\sl S. marcescens} dense populations, even if the initial ratio between species was in favor of the {\sl B. subtilis}. In these cases, we find that {\sl B. subtilis} are migrating through the {\sl S. marcescens} and take advantage of almost the entire volume of the colony (Fig. 2d). In particular, they do not form a thin layer as in axenic {\sl B. subtilis} colonies. At very low initial concentrations of {\sl S. marcescens}, the colony is flat with only a few layers of cells. The coexistence of the two species may suggest mutualistic interaction, which enhances cell growth in the presence of other species, as obtained in our liquid assays and also reported previously \cite{ref31,ref43,ref44,ref45}. 
Mixed colonies can form either thick or thin structures, and their interface growth depends monotonically on the initial ratio between species (Fig. 3f). Thinner colonies formed when the initial ratio of cells was in favor of {\sl B. subtilis}, probably due to the stronger surfactant activity of the {\sl B. subtilis} (surfactin vs. serrawettin) that reduces the surface tension of the entire colony \cite{ref21,ref34}. This caused the {\sl B. subtilis} to migrate faster, forming two apparent fronts (Fig. 1). With that, the {\sl S. marcescens} benefit in the mix as the fast-expanding {\sl B. subtilis} enhances their expansion speed (with respect to axenic {\sl S. marcescens}). Chemotaxis assays have yielded negative results suggesting that the mixed colonies do not exhibit chemotactic interactions. Accordingly, there was no local phase separation on the microscopic scale, for example into distinct clusters, showing that there is no antagonistic interaction between them. Contrarily, the more the species were mixed, the faster they moved (Fig. 3c), exhibiting larger correlation lengths (Fig. 3e) in the velocity field and a Gaussian distribution (Fig. 3c). 

Overall, our results are unique in the sense that they show quantitative microscopic interactions between two swarm species mixed jointly in the same colony. Moreover, it demonstrates advantageous interactions between the species across a range of scales.

\section*{Acknowledgments} 
We thank Rasika M. Harshey and Daniel B. Kearns for sending the strains and Avigdor Eldar for creating the fluorescent variants. Partial support from the Deutsche Forschungsgemeinschaft (The German Research Foundation DFG) Grants No. HE5995/3-1and No. BA1222/7-1 is thankfully acknowledged.

\clearpage
\newpage

\begin{figure}[h!tp]
\centering
\includegraphics[trim={0cm 22cm 10cm 0cm},clip, width=15cm, angle=0]{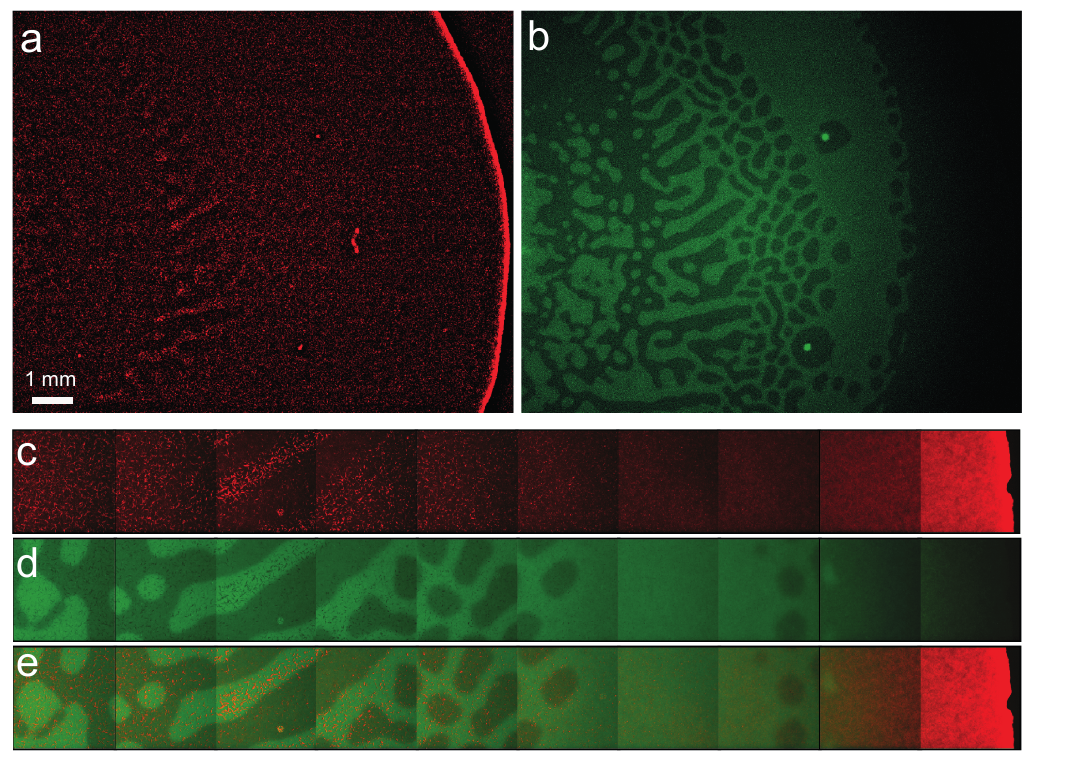}
\caption{
A top-view macroscopic image of a mixed swarm colony consisting of {\sl B. subtilis} (red) and {\sl S. marcescens} (green). (a) the {\sl B. subtilis} population, and (b) the {\sl S. marcescens} population. (c)-(e) higher magnification at increasing distances from the colony center: (c) {\sl B. subtilis}, (d) {\sl S. marcescens}, and (e) the superimposed image of (c) and (d). The initial ratio between the species is 75\% {\sl B. subtilis} and 25\% {\sl S. marcescens}.  }
\label{fig1}
\end{figure}
\begin{figure}[h!tp]
\centering
\includegraphics[trim={0cm 20cm 10cm 0cm},clip, width=15cm, angle=0]{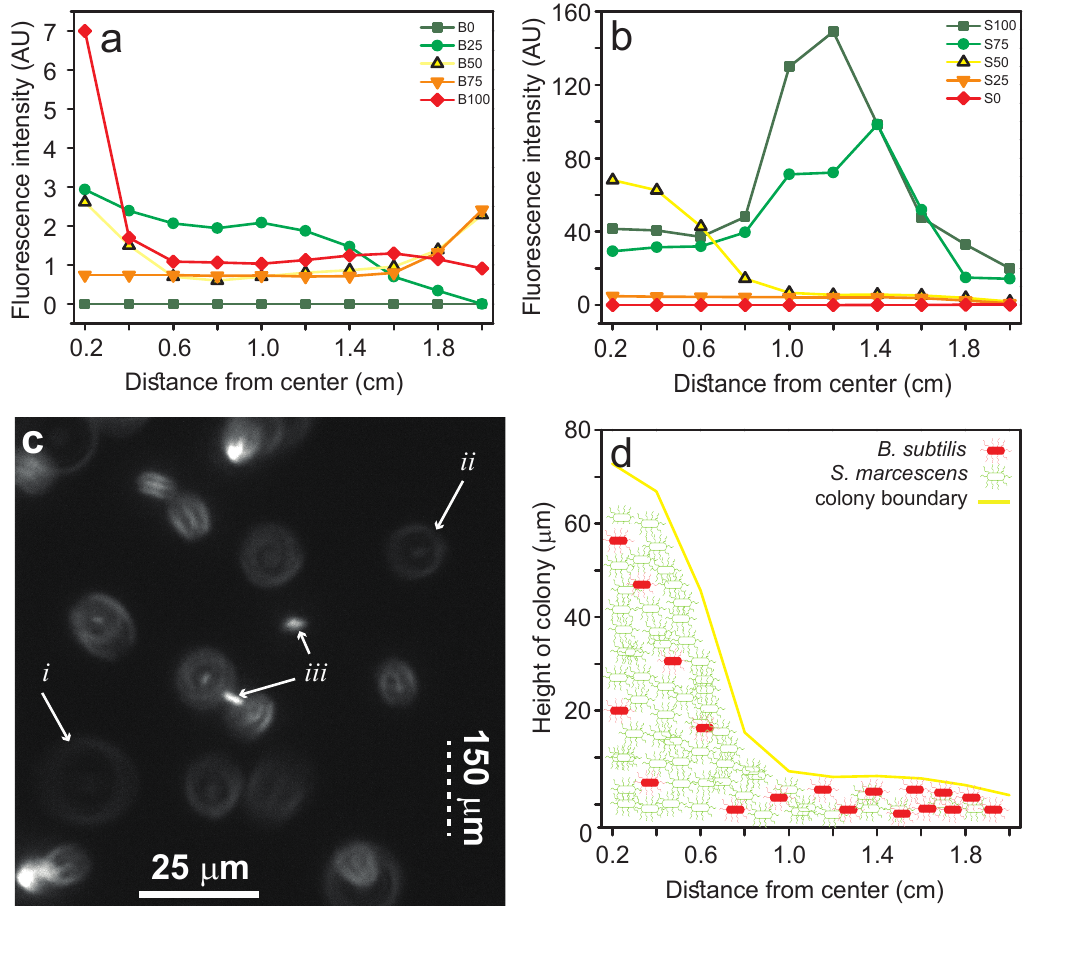}
\caption{
 The three-dimensional (3D) distribution of species in a mixed colony. (a) The average fluorescence intensity (indicative of the height of the colony) of {\sl B. subtilis} for different initial mixing ratios as a function of the distance from center. Each value on the perpendicular axis corresponds to a single layer of cells. The value is computed from a top-view so the cells do not necessarily occupy a fixed layer. (b) Same as (a) for {\sl S. marcescens}. (c) A snapshot from an off-focus fluorescence mode showing the diffraction rings of the {\sl B. subtilis} cells embedded in a {\sl S. marcescens} colony. Arrows point to rings with a variety of diameters, reflecting the locations of the cells on the depth ($z$-axis) of the 3D colony. Solid scale-bar indicates lateral dimension. Dashed scale-bar indicates the depth. The mixing ratio is 50\% {\sl S. marcescens} and 50\% {\sl B. subtilis} (where only 10\% are labeled). (d) Cross sectional schematic representation of a mixed colony, illustrating the approximate 3D structure and dissemination of each species in the mixed colony. }
\label{fig2}
\end{figure}
\begin{figure}[h!tp]
\centering
\includegraphics[trim={0cm 16cm 10cm 0cm},clip, width=15cm, angle=0]{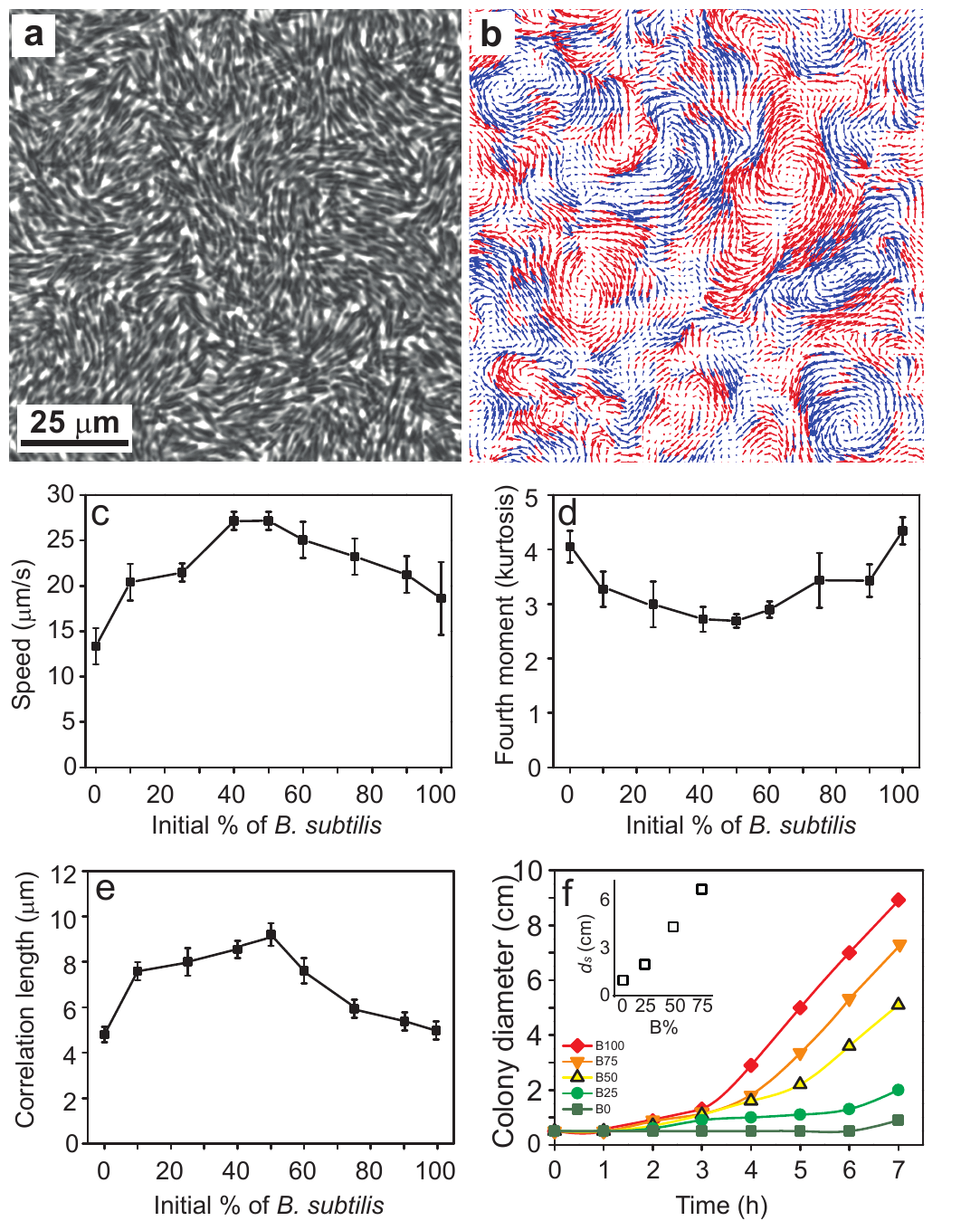}
\caption{
The swarming dynamics of a mixed colony. (a) A top-view phase contrast image of the active area in a mixed swarm. Image was taken approximately 100 $\upmu$m away from the edge of a 4 cm diameter colony. (b) The velocity field of the same image shown in (a). Arrows show the instantaneous direction of motion. Blue indicates CCW motion and red indicates CW motion. (c)-(e) Experimental results obtained from the collective dynamics of a mixed colony as function of the initial concentration of {\sl B. subtilis} in the mix. (c) The average speed, (d) the kurtosis of the velocity distribution, and (e) The correlation length. (f) The macroscopic expansion rate of each mixed colony, for a variety of initial mixing concentrations. Inset shows the diameter of the inner {\sl S. marcescens} front at $t=7$ h, $d_s$, for several initial fractions of {\sl B. subtilis}.   }
\label{fig3}
\end{figure}
\begin{figure}[h!tp]
\centering
\includegraphics[trim={0cm 19cm 10cm 0cm},clip, width=15cm, angle=0]{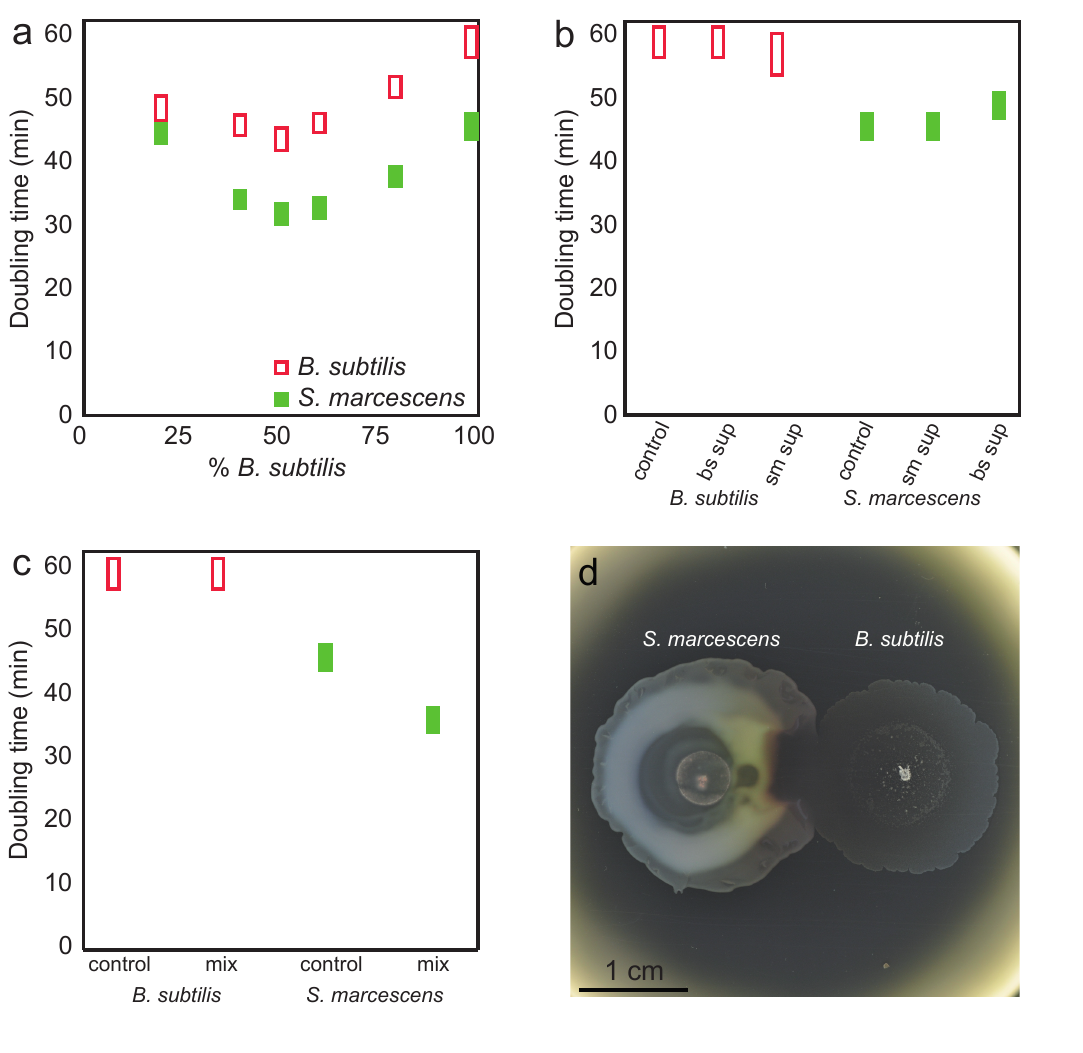}
\caption{
Interspecies interaction in bulk liquid: (a) Doubling time of each species when cocultured in liquid medium as function of their initial concentration. (b) Doubling time of each species in different supernatants. LB broth is used as control. (c) Doubling time of each species in U-tube chamber. For control, the same species is inoculated in each side of the U-tube chamber. (d) Chemotaxis assay; growing two axenic colonies in a short proximity. This image was taken several hours after inoculation; {\sl S. marcescens} was inoculated a few hours prior to {\sl B. subtilis} in order to obtain colonies that are similar in size. No chemotaxis was observed if the two colonies were inoculated at the same time. No attraction, repulsion or directional bias was observed. The same result was obtained for colonies inoculated at the same time.}
\label{fig4}
\end{figure}

\end{document}